\documentclass[twoside,12pt]{article}
\usepackage{amssymb,amsmath}
\def\R{\Bbb{R}}
\def\N{\Bbb{N}}
\def\M{\Bbb{M}}

\def\Box{\raisebox{3pt}{\fbox{}}\hspace{2pt}}
\def\supp{{\mbox{\rm supp}\, }}
\def\C{\Bbb{C}}

\newtheorem{theorem}{Theorem}[section]
\newtheorem{condition}[theorem]{Condition}

\newtheorem{corollary}[theorem]{Corollary}

\newtheorem{lemma}[theorem]{Lemma}

\begin{document}
\pagestyle{myheadings}
\markboth{H. Gottschalk, H. Thaler}{Interacting QFT with background gravitation}
 \thispagestyle{empty}
\begin{center}
{\Large \bf An indefinite metric model for interacting quantum fields with non-stationary background gravitation}
\end{center}

\

\begin{center} {\sc Hanno Gottschalk} and {\sc Horst Thaler}\\~\\
{\small Institut f\"ur angewandte Mathematik, \\ Rheinische
Friedrich-Wilhelms-Universit\"at,\\ Wegelerstr. 6, D-53115
Bonn, Germany\\ e-mail: gottscha@wiener.iam.uni-bonn.de\\
horst@wiener.iam.uni-bonn.de}
\end{center}

\
\begin{center} 30.01.2002\end{center}
\

{\noindent \small {\bf Abstract.}  We consider a relativistic {\em ansatz} for the vacuum expectation values (VEVs) of a quantum field on a globally hyperbolic space-time
 which is motivated by certain Euclidean field theories. The Yang-Feldman asymptotic condition w.r.t. a "in"-field in a quasi-free representation of the canonic commutation relations (CCR) leads
to a solution of this {\em ansatz} for the VEVs.  A GNS-like construction on a non-degenerate inner product space then gives local, covariant quantum fields
with indefinite metric on a globally hyperbolic space-time. The non-trivial scattering behavior of quantum fields is analyzed by
construction of the "out"-fields and calculation of the scattering matrix. A new combined effect of non-trivial quantum scattering and
non-stationary gravitational forces is described for this model, as quasi-free "in"- fields are scattered to "out"-fields which form a
non quasi-free representations of the CCR. The asymptotic condition, on which the construction is based, is verified for the concrete example
of de Sitter space-time. }

\vspace{.25cm}
{\small \noindent {\bf Keywords}: {\it Quantum fields on curved space-time, scattering, {\rm "in"}- and {\rm "out"}-
representations, spectral condition, de Sitter space-time.}}
 %\tableofcontents
 \section{Introduction}
 \label{1sect}
 The interest in quantum fields in curved space-times stems from the very 
 physical question how a curved geometry combines with quantum effects.  The particle 
 production observed in the case of non-stationary background gravitation (related to the Hawking effect \cite{Haw,Wa})
 is possible since the state of the system restricted to different space time regions gives 
 in general rise to different representations which account for the 
 particle production. In the present work we investigate a similar effect for a class of interacting 
 indefinite metric quantum field models on globally hyperbolic space-times. 
 
 The models, originating from Euclidean quantum
  field theory (QFT) using Poisson fields \cite{AGW1,AGW2}, in the case of Minkowski space-time give rise to solutions to the 
 modified Wightman axioms of Morchio and Strocchi \cite{MS} which exist in space-time
 dimensions higher than three and show a non-trivial scattering behavior, see \cite{AG,AGW1}. It turns out that a similar construction
 can be carried out on quite general globally hyperbolic manifolds (Sections 2 and 3) without directly deriving them from Euclidean models, which however remain
  a source of "inspiration". The theoies' vacuum expectation values fulfill the requirements of locality, invariance under time-orientation preserving isomorphisms and Hermiticity (Section 4). 
They thus give rise to a GNS-like representation of the field algebra on an non-degenerate inner product space and hence an "indefinite metric" QFT (see Appendix A). The restriction of this
 representation to the sub-algebras generated by incoming and outgoing fields can then be investigated through algebraic methods.      
   
 Our emphasis is laid upon a scattering theory based on Yang-Feldman equations \cite{YF} which also unveils the way how a curved geometry 
 influences the character of a free field when it travels through space-time and at the same time undergoes a quantum-mechanical scattering.  
 Preparing the state for the algebra of 
 "in"-fields in a quasi free representation, existence of "out"-fields
 fields hinges upon the dispersion properties of the fundamental solutions of the Klein-Gordon equation and necessitates the formulation of a so called 
 dispersion condition whose fulfillment may depend not only on the manifold structure but also on 
 the dimension. 
 
 This allows the calculation of the matrix elements between "in"- and "out"-fields which, as in the flat case \cite{AG}, expose non-trivial
 scattering.  
 When observing the matrix elements of only the "out"-fields it turns out that they describe 
 a representation of the CCR which in general is rather different from that of the "in"-fields (Section 5) . In particular,
 "in"-fields in quasi-free representations are scattered to "out"-fields in non quasi-free representations.
 What leads to this effect is a combination of non-trivial quantum scattering and time-dependent 
 gravitational forces. It can not be reduced to the conceptually similar, but mathematically different change of representations
 related to the Hawking-effect \cite{Haw,Wa}.  
 "In"- and "out"-representations are shown to be equivalent in the case of stationary space-times, where a spectrum condition can be 
 formulated (Section 6). 
 
 As a concrete example, we verify the dispersion condition for de Sitter space-time (Section 7).
 
 The scope of this paper is to communicate observations made in the case of our comparatively simple model. At this point
 it is natural to put forward the question, whether the observed effects also play a r\^ole in the case of more physically motivated 
 perturbative constructions \cite{BrFr,Tag} of interacting quantum fields on non stationary globally hyperbolic space-times. Our conjectured answer
 to this question is "yes". The findings of this paper can be related with effects in first order $:\phi^n:$-perturbation theory 
 on a non-stationary globally hyperbolic manifold using the calculus of sectorized Feynman graphs of A. Ostendorf \cite{Os} and O. Steinmann. \cite{Ste1,Ste2}.
In fact, up to different initial conditions (the Feynman rules of \cite{Os,Ste1,Ste2} on a non-stationary space-time manifold do not lead to a quasi-fee "in"- or "out"-state)
our $n$-point functions coincide with the evaluation of the "star"-graph with one vertex and $n$ legs, which is the
first order contribution to the connected $n$-point function, see  \cite{GT} for a detailed analysis. This indicates that non-quasi-free representations
of the CCR have a natural place in interacting QFT on curved space-time and might enhance the recent interest in this topic \cite{HR}.

\section{The relativistic {\it ansatz}}
\label{3sect}
 We want to construct Wightman functions on a $d$-dimensional Lorentzian manifold\footnote{\label{2.1foot}i.e. the metric $g$ carries signature $(1,-1,\ldots,-1)$.} $(\M,g)$ which is globally
 hyperbolic with distinguished time direction\footnote{\label{2.2foot}Here we interpret this notion in the restrictive sense that as a $C^\infty$ manifold $\M\cong\R\times\Sigma$ with $\Sigma$ a $d-1$ dimensional manifold and a time direction given by the relation $<$ on $\R$.} but not necessarily stationary\footnote{\label{2.3foot}In the coordinates $(t,\vec x)\in\R\times\Sigma$ the metric tensor $g$  
 in general depends on $t$ and $\vec x$.}. For $x,y\in \M$, $x\not=y$ we say that $x$ and $y$, $x\not=y$, are light-like/time-like/space-like separated, if they can be
 connected by a light-like/time-like curve\footnote{\label{2.4foot}By a curve $c:[0,1]\to \M$ s.t. $g(c'(s),c'(s))=0$ / $>0$ $\forall s\in[0,1]$.} or if there is no such curve, respectively. That $x$ is space-like to $y$ is expressed by the symbol $x\perp y$. For $x$ and $y$ light-like or time-like separated and $x$ later than $y$ (in the time-direction fixed on $\M$) 
 we write $x\geq y$. By $V^\pm_x$ we denote the open forward/backward light-cone with base-point $x$, i.e. $\bar V^\pm_x=\{y\in \M:y\geq (\leq)x\}\cup\{x\}$. 
 
 We next introduce the fundamental functions following \cite{Di}. Let $G_{\rm r}$ and $G_{\rm a}$ be the retarded / advanced fundamental solution of the d'Alambertian $(\Box+m^2)$, i.e. $G_{\rm r/a}$ are real-valued
 distributions such that ${\rm supp}_y~G_{\rm r/a}(f,y)\subseteq \cup_{x\in\supp f} \bar V_{x}^\mp$, $f\in {\cal D}(\M)=C_0^\infty(\M)$, and
\begin{equation}
\label{2.1eqa}
G_{\rm r/a}((\Box+m^2)f,h)=G_{\rm r/a}(f,(\Box+m^2)h)=\int_\M fh\, dx\, ,~~~f,h\in {\cal D}(\M),
\end{equation} 
with $dx$ the canonic volume form associated with $g$. These conditions determine $G_{\rm r/a}$ uniquely \cite{Di}.
We also note that $G_{\rm r}(f,h)=G_{\rm a}(h,f)$. Next we define the antisymmetric commutator function $D(f,h)$ as 
 \begin{equation}
 \label{2.2eqa}
 D(f,h)=G_{\rm r}(f,h)-G_{\rm a}(f,h)\,.
 \end{equation}
 Obviously, $D$ fulfills the Klein-Gordon equation in both arguments, i.e. 
 \begin{equation}
 \label{2.2aeqa}
 D((\Box+m^2)f,h)=D(f,(\Box+m^2)h)=0.
 \end{equation}
  Furthermore, $D(f,h)=0$ for $\supp f\perp \supp h$ as a consequence of the
 support properties of $G_{\rm r/a}$.
 
 Let $D^+$ be a (complex valued) distribution in ${\cal D}(\M\times \M,\C)'$ such that ${\rm Im}D^+=D$, 
 \begin{equation} 
 \label{2.3eqa}
 D^+((\Box+m^2)f,h)=D^+(f,(\Box+m^2)h)=0
 \end{equation}
 and furthermore $D^-(f,h)=D^+(h,f)=\overline{D^+(\bar f,\bar h)}$, i.e. the real part of $D^+$ is symmetric. We furthermore demand that $D^+$ is invariant under isometric diffeomorphisms preserving the time orientation,
 $D^+(f_\alpha,h_\alpha)=D^+(f,h)$, $\alpha\in{\cal G}^\uparrow(\M,g)$, $f_\alpha(x)=f(\alpha^{-1}(x))$. As this property automatically is fulfilled for the imaginary part $D$ of $D^+$,
 this is only a condition on the real part of $D^+$.  Lastly, we demand that $D^+(f,y)$ is a measurable function in $y$ and $D^+(\bar f,f)\geq 0$ $\forall f\in {\cal D}(\M)$. For a discussion on the existence of a (not necessarily ${\cal G}^\uparrow(\M,g)$-invariant) $D^+$ see \cite[Chapter 4.2]{Wa} -- for
 $D^+$ with the Hadamard property, $D^+(f,y)$ is a measurable function in $y$ $\forall f\in{\cal D}(\M,\C)$, cf. \cite[Chapter 4.6]{Wa}. Lastly, the ${\cal G}^\uparrow(\M,g)$ invariance of at least one such Hadamard state can be justified from the physical belief that
 a "good" vacuum state has maximal symmetry properties. This is e.g. true for Minkowski- and de Sitter space-time (see Section \ref{8sect}) and for stationary space-times the uniqueness of states fulfilling the spectral condition implies at least invariance under time-translations (see Section \ref{7sect}).  
 
 Under these conditions, the two-point function $D^+$, via second quantization, 
gives rise to a ${\cal G}^\uparrow(\M,g)$-covariant representation of the canonic commutation relations (CCR) on a Hilbert space ${\cal H}$ through free fields,  cf. (\ref{3.2eqa})--(\ref{3.4eqa}) below. It is a well-known and fundamental
fact for quantum field theory that in the absence of the spectral property and for a possibly "too small" symmetry group ${\cal G}^\uparrow(\M,g)$ the real part of $D^+$ is not uniquely fixed and thus on curved space-time there are many non-equivalent
admissible representations of the CCR \cite{Wa}.

We now write down the equations for the truncated Wightman functions $\langle\Psi_0,\phi(f_1)\cdots\phi(f_n)\Psi_0\rangle^T$ of our model. The Wightman functions, or vacuum expectation values, are
then given in terms of truncated Wightman functions via   
\begin{equation}
\label{2.4eqa}
\left\langle\Psi_0,\phi(f_1)\cdots\phi( f_n)\Psi_0\right\rangle=\sum_{I\in{\cal P}^{(n)}}\prod_{\{j_1,\ldots j_l\}\in I}\left\langle\Psi_0,\phi(f_{j_1})\cdots  \phi(f_{j_l})\Psi_0\right\rangle^T
\end{equation}   
where ${\cal P}^{(n)}$ is the collection of all partitions of $\{1,\ldots,n\}$ into disjoint, nonempty subsets $\{j_1,\ldots,j_l\}$ where
$j_1<\ldots<j_l$. A comment concerning the use of symbols like $\langle\Psi_0,\phi(f_1)\cdots\phi(f_n)\Psi_0\rangle$ is in order. Until Section 5 they only stand for distributions in ${\cal D}(\M^{\times n},\C)'$. Once the
Hermiticity condition is verified for these distributions in Sections 4 and 5, one can apply the GNS-construction on inner product spaces, see Appendix A, and one {\em a posteriori} verifies that $\Psi_0$ has a proper mathematical meaning as the GNS-vacuum,
$\phi(f)$ (and also the incoming and outgoing fields $\phi^{\rm in/out}(f)$) as operator valued distributions on the non-degenerate inner product space and $\langle\Psi_0,\phi(f_1)\cdots\phi(f_n)\Psi_0\rangle$ as vacuum expectation values w.r.t.
 the non-degenerate inner product $\langle.,.\rangle$. Anticipating this standard construction,
we use this notation from the beginning.

Let $\langle\Psi_0,\Psi_0\rangle=1$, $\langle\Psi_0,\phi(f)\Psi_0\rangle^T=0$  and 
\begin{equation}
\label{2.5eqa}
\left\langle\Psi_0,\phi(f)\phi(h)\Psi_0\right\rangle^T={b^2\over m^2}D^+(f,h)~~~~\forall f,h\in{\cal D}(\M,\C)
\end{equation}
for some $b>0$. For the higher order truncated Wightman functions containing one "current" entry $j(f)=\phi((\Box+m^2)f)$ we set 
\begin{eqnarray}
\label{2.6eqa}
&&\left \langle\Psi_0,\phi(f_1)\cdots\phi(f_{k-1})j(f_k)\phi(f_{k+1})\cdots\phi(f_n)\Psi_0\right\rangle^T~~~~~~~~~~~~~~~~~\nonumber\\
&&~~~=b_n\prod_{l=1}^{k-1}D^-(f_l,f_k)\prod_{l=k+1}^nD^+(f_l,f_k),~~f_l\in{\cal D}(\M,\C),~l=1,\ldots,n.
\end{eqnarray}
Again, $b_n\in \R$ is some arbitrary constant. 
In the next step we fix the Wightman function with two (or more) "current" entries
\begin{equation}
\label{2.7eqa}
\left \langle\Psi_0,\phi(f_1)\cdots\phi(f_{k-1})j(f_k)\phi(f_{k+1})\cdots\phi(f_{r-1})j(f_r)\phi(f_{r+1})\cdots\phi(f_n)\Psi_0\right\rangle^T=0,
\end{equation}
i.e. any truncated vacuum expectation value containing two (or more) current operators $j$ vanishes. 

Before solving the above {\em ansatz} by specifying initial conditions in the next section, we would like to briefly sketch 
from Euclidean QFT which motivates the Equations (\ref{2.5eqa}) -- (\ref{2.7eqa}). We suppose that there exists a Riemannian manifold 
obtained as the analytic continuation of $\M$ to purely imaginary Euclidean time. We consider an Euclidean field theory on the Riemannian manifold
 which is 
the solution of the stochastic partial differential equation ($\Delta$ denotes the Laplacian)  
\begin{equation}
\label{2.8eqa}
-\Delta\varphi+m^2\varphi=\eta\,.
\end{equation}    
where $\eta$ is some noise field with a mixed Gauss-Poisson distribution \cite{AGW1}. The noise field $\eta$
can thus be seen as the Euclidean analogue of the current $j$. A proper choice of $\eta$ \cite{AGW2} then leads to 
a solution $\varphi$ which in the flat case has Schwinger (moment) functions which can be analytically continued
to real relativistic time \cite{AGW1}. The solutions in this case coincide with the solution given in the next section.

To be just a little bit more detailed, let us emphasize that Equation (\ref{2.7eqa}) on the Euclidean side is entailed  
by the fact that truncated correlations of noise fields vanish at the non-coinciding points - and only such points would
matter for a {\em bona fide} analytic continuation. The remaining two equations, (\ref{2.5eqa}) and (\ref{2.6eqa}) then can
be directly traced back to the analytic continuation of Euclidean two-point functions of the random field model. If $D^\pm$ are these
analytic continuation depending on the Euclidean time-ordering, (\ref{2.6eqa}) is the relativistic analogue of a related Euclidean
equation describing the coupling of $\varphi$ to the noise-field $\eta$. The constants $b$ and $b_n$ depend on the probability distribution of
$\eta$, in particlular $b_n=0$, $n\geq 3$, if $\eta$ is purely Gaussian.

\section{Construction of Wightman functions}
\label{4sect}
If one could uniquely invert the operator $\Box+m^2$, the equation (\ref{2.6eqa}) would uniquely determine the truncated vacuum expectation values of the field $\phi$, however this is not the case.
To get the vacuum expectation value with $\phi(f_k)$ instead of $j(f_k)=\phi((\Box+m^2)f_k)$ it is necessary to specify the initial conditions for the field $\phi(x)$. Our choice to do this is to specify initial conditions
for large times $x^0\to\mp\infty$ and to postulate that for such asymptotic times the local field $\phi(x)$ converges to free incoming or outgoing fields $\phi^{\rm in/out}(x)$. The adequate technical formulation is 
given by the Yang-Feldman equations \cite{YF}
\begin{eqnarray}
\label{3.1eqa}
&&\langle\Psi_0,\phi(f_1)\cdots \phi(f_{k-1})\phi(f_k)\phi(f_{k+1})\cdots\phi(f_n)\Psi_0\rangle^T\nonumber\\
&=&\langle\Psi_0,\phi(f_1)\cdots \phi(f_{k-1})\phi^{\rm in/out}(f_k)\phi(f_{k+1})\cdots\phi(f_n)\Psi_0\rangle^T\nonumber\\
&+&\langle\Psi_0,\phi(f_1)\cdots \phi(f_{k-1})j(G_{\rm r/a}f_k)\phi(f_{k+1})\cdots\phi(f_n)\Psi_0\rangle^T\, ,
\end{eqnarray}  
for $k=1,\ldots, n$ and $n\in \N$, $f_l\in{\cal D}(\M,\C)$, $G_{{\rm r/a}}f_k(y)=G_{{\rm r/a}}(f_k,y)$. Clearly, $G_{{\rm r/a}}f_k\not\in {\cal D}(\M,\C)$ for $f_k\not=0$ which can be seen as an "infra-red" problem, cf. 
the discussion preceding Condition \ref{3.1cond} below.

From (\ref{2.4eqa})--(\ref{2.5eqa}) and (\ref{3.1eqa}) we immediately get that the choice
\begin{equation}
\label{3.2eqa}
\left\langle\Psi_0,\phi^{a}(f)\phi^{b}(h)\Psi_0\right\rangle^T={b^2\over m^2}D^{+}(f,h) ~~\mbox{ for }a,b={\rm in/loc/out}
\end{equation}
for the truncated two point function is a uniquely given solution which agrees with the postulates which have been set up so far. 
Let us next consider the problem for the truncated $n$-point functions, $n\geq 3$.

Here it is also important to note that one can only fix the initial or the final behavior of $\phi$, but not both, as this would over-determine the system.
Here we use the conventions that (\ref{3.1eqa}) thus for the in-case is part of the {\it ansatz} whereas for the out-case it is a statement which we have to verify
for the solution we give.

Likewise, we have to postulate the characteristic properties of a free field for $\phi^{\rm in}$, namely
that free fields fulfill the CCR and the Klein-Gordon equation. But we also have to specify a {\em representation} for the
in-fields, as in the absence of the spectral condition invariance, equations of motion and commutation relations do suffice to fix the
representation uniquely. Our choice for the representation of the in-field is (\ref{3.2eqa}) for the two point function and
\begin{equation}
\label{3.3eqa}
\left\langle\Psi_0,\phi^{\rm in}(f_1)\cdots\phi^{\rm in}(f_n)\Psi_0\right\rangle^T=0~~\mbox{ for }n\geq 3,
\end{equation}
hence we want the "in"-field to be in a quasi-free representation \cite{Wa}. Once the GNS-like construction of Appendix \ref{Aapp} has been carried through and $\phi^{\rm in}(f)$ 
is realized as an operator valued distribution on the indefinite metric state space, (\ref{3.3eqa}) together with (\ref{3.2eqa}) immediately implies
\begin{equation}
\label{3.4eqa}
\phi^{\rm in}((\Box+m^2)f)=0,~~~[\phi^{\rm in}(f),\phi^{\rm in}(h)]=i{b^2\over m^2}D(f,h)
\end{equation}
whereas the analogous statement for $\phi^{\rm out}$ have to be proven\footnote{By a simple re-definition of field strengths $\phi\to (m/b)\phi$ one obtains the usual normalization of the CCR.}.

Before we proceed on the basis of the above assumptions, we have to discuss whether the asymptotic condition (\ref{3.1eqa}) makes sense at all. 
The existence of free asymptotic fields can not be expected for an arbitrary Lorentzian manifolds $\M$. If we 
e.g. consider a stationary Lorentzian manifold $\M=\R\times\Sigma$ with $g$ depending only on $\vec x$ and $\Sigma$ compact (cf. footnotes \ref{2.2foot},\ref{2.3foot}) then we have neither a
natural dispersion of wave-packets in non-compact space (for sufficiently high dimension $d$ of the space-time $\M$) nor a dispersion which is due to the expansion of the space-time at asymptotic times. We therefore 
need a criterion on our manifold $(\M,g)$ which implies that either (or both) of the above dispersion effects is strong enough to guarantee the asymptotic condition (\ref{3.1eqa}) with $\phi^{\rm in/out}$ free fields. 
Such dispersion is most conveniently formulated in terms of the fundamental function $D^+$ which determines all the other fundamental functions:
\begin{condition}
\label{3.1cond}
(i) On the manifold $(\M,g)$ there exists a $D^+$ as described in Section \ref{3sect} that fulfills the $n$-point dispersion condition $ \int_\M|D^+(f,x)|^n\,dx$
$<\infty$ $\forall f\in {\cal D}(\M,\C)$ for $n\geq 3$ or $n\geq 4$.

\noindent (ii) Furthermore, if $f_l\to 0$ in ${\cal D}(\M,\C)$ then $D^+(f_l,x)\to 0$ $\forall x\in \M$ and $\exists$ a measurable function $F(x)\geq |D^+(f_l,x)|$ $\forall x\in \M,l\in\N$ s.t.
$\int_{\M}F(x)|D^+(f,x)|^{n-1}\, dx<\infty,\;\forall f \in {\cal D}(\M,\C)$.
\end{condition}

It is clear that the condition for $D^+(f,x)$ that one or several functions $D^+(f,x)$ can be replaced by other fundamental functions $D^-(f,x),D(f,x)$ and
also \linebreak $G_{\rm r/a}(f,x)$ as the latter outside of the causal closure of the compact support of $f$ behave like $D^+(f,x)$ or are equal to zero. 
For $4$-dimensional Minkowski space-time Condition \ref{3.1cond} ($n=3$) follows from the decay behavior of fundamental functions in arbitrary directions, see e.g. \cite{Ru}, and for de Sitter space-time we will verify Condition \ref{3.1cond} in Section \ref{8sect}.

Let us thus resume the construction of Wightman functions for $n\geq 3$. By the Yang-Feldman equations for the in-case, we can replace one local field by a "in"-field and a current, e.g. $\phi(f_1)$
\begin{eqnarray}
\label{3.5eqa}
\left\langle\Psi_0,\phi(f_1)\phi^{\rm in}(f_2)\cdots\phi^{\rm in}(f_n)\Psi_0\right\rangle^T&=&
\left\langle\Psi_0,\phi^{\rm in}(f_1)\cdots\phi^{\rm in}(f_n)\Psi_0\right\rangle^T\nonumber\\
&+&\left\langle\Psi_0,j(G_{\rm r}f_1)\phi^{\rm in}(f_2)\cdots\phi^{\rm in}(f_n)\Psi_0\right\rangle^T\nonumber\\
&=&b_n\int_\M G_{\rm r}(f_1,y)\prod_{l=2}^nD^+(f_l,y)\, dy\, .
\end{eqnarray}
Here we used (\ref{3.3eqa}) and we exploited the fact that by the Yang--Feldman equations (\ref{3.1eqa}) the difference between the local and the "in"-field 
is given by a current in order to replace "in"-fields by local fields according to the equations (\ref{2.7eqa}) and we finally evaluated the vacuum expectation value 
containing one current and local fields by (\ref{2.6eqa}). By Condition \ref{3.1cond} the integral of the right hand side of (\ref{3.5eqa}) converges. Also, the right hand side of (\ref{3.5eqa}) is a distribution in ${\cal D}(\M^{\times n},\C)'$.
To see this, we let one of the test functions $f_l$ go to zero in ${\cal D}(\M,\C)$. From Condition \ref{3.1cond} (ii) it then follows that the right hand side
of (\ref{3.5eqa}) converges to zero by Lebesgue's theorem of dominated convergence.
   
By induction, we can now calculate 
$\langle\Psi_0,\phi(f_1)\cdots\phi(f_k)\phi^{\rm in}(f_{k+1})\cdots\phi^{\rm in}(f_n)\Psi_0\rangle^T$ using (\ref{3.5eqa}) and the same arguments as in the first step. Continuing in this way, we get after $k=n$ steps 
the Wightman functions of the local fields:
\begin{equation}
\label{3.6eqa}
\left\langle\Psi_0,\phi(f_1)\cdots\phi(f_n)\Psi_0\right\rangle^T=b_n\sum_{k=1}^n\int_\M\prod_{l=1}^{k-1}D^-(f_l,y)G_{\rm r}(f_k,y)\prod_{l=k+1}^nD^+(f_l,y)\, dy\, ,
\end{equation}
where again Condition \ref{3.1cond} assures that the integrals in (\ref{3.6eqa}) exist and define a distribution in ${\cal D}(\M^{\times n},\C)'$.  It also does not matter in which order we replace the $\phi^{\rm in}$ in 
(\ref{3.3eqa}) by local fields, as this only changes the order of summation in (\ref{3.6eqa}). We have thus derived
\begin{theorem} 
\label{3.1theo} 
Let Condition \ref{3.1cond} be fulfilled. Given the {\it ansatz} (\ref{2.5eqa}), (\ref{2.6eqa}), (\ref{2.7eqa}) and the asymptotic condition (\ref{3.1eqa}) for a free 
 "in"-field in a quasi-free representation (cf. (\ref{3.3eqa})--(\ref{3.4eqa})), there exists a unique solution for the (truncated) Wightman functions of the local field given by (\ref{3.2eqa}) and (\ref{3.6eqa}).
\end{theorem}

\section{Properties of Wightman functions}
 \label{5sect}
In this section we verify the fundamental properties of the Wightman functions
 constructed in Section \ref{4sect}.

\begin{theorem}
\label{4.1theo}
The Wightman functions constructed in Section \ref{4sect} fulfill the properties of Hermiticity, ${\cal G}^\uparrow(\M,g)$-invariance and
locality.
\end{theorem}

 We start with the proof of the first property, Hermiticity:
\begin{equation}
\label{4.1eqa}
\left\langle\Psi_0,\phi(f_1)\cdots\phi(f_n)\Psi_0\right\rangle^T=\overline{\left\langle\Psi_0,\phi(\bar f_n)\cdots\phi(\bar f_1)\Psi_0\right\rangle^T}\, .
\end{equation} 
For $n=2$ (\ref{4.1eqa}) follows from the properties of $D^+$. For larger $n\in\N$, this relation can be easily verified from (\ref{3.6eqa}) as complex conjugation of the right hand side exchanges $D^\pm(\bar f_l,y)$ with $D^\mp(f_l,y)$ and $G_{\rm r}(\bar f_k,y)$ with $G_{\rm r}(f_k,y)$. After 
re-ordering the sum, we then see that the complex conjugate is just the r.h.s. of (\ref{3.6eqa}) with the reversed order of the arguments.

Let next $\alpha\in{\cal G}^\uparrow(\M,g)$. To verify invariance, we first note that $G_{\rm r}(f_\alpha,h_\alpha)=\linebreak G_{\rm r}(f,h)$ and analogous identities hold
for $D^\pm$. Hence,
\begin{eqnarray}
\label{4.2eqa}
&&\left\langle\Psi_0,\phi(f_{1,\alpha})\cdots\phi(f_{n,\alpha})\Psi_0\right\rangle^T\nonumber\\
&=&b_n\sum_{k=1}^n\int_\M\prod_{l=1}^{k-1}D^-(f_{l,\alpha},y)G_{\rm r}(f_{k,\alpha},y)\prod_{l=k+1}^nD^+(f_{l,\alpha},y)\, dy\nonumber\\
&=&b_n\sum_{k=1}^n\int_\M\prod_{l=1}^{k-1}D^-(f_l,\alpha^{-1}(y))
G_{\rm r}(f_k,\alpha^{-1}(y))\prod_{l=k+1}^nD^+(f_l,\alpha^{-1}(y))\, dy
\end{eqnarray}
As $dy$ is $\alpha^{-1}$-invariant, the r.h.s. coincides with the right hand side of (\ref{3.6eqa}) which establishes invariance under ${\cal G}^\uparrow(\M,g)$.

It remains to verify locality. By explicit calculation we get
\begin{eqnarray}
\label{4.3eqa}
&&\left\langle\Psi_0,\phi(f_1)\cdots[\phi(f_k),\phi(f_{k+1})]\cdots\phi(f_n)\Psi_0\right\rangle^T\nonumber\\
&=&b_n\int_\M\prod_{l=1}^{k-1}D^-(f_l,y)[ G_{\rm r}(f_k,y)D^+(f_{k+1},y)+D^-(f_k,y)G_{\rm r}(f_{k+1},y)\nonumber\\
&-&G_{r}(f_{k+1},y)D^+(f_k,y)-D^-(f_{k+1},y)G_{\rm r}(f_k,y)]\prod_{l=k+2}^nD^+(f_k,y)\, dy\nonumber\\
&=&ib_n\int_\M\prod_{l=1}^{k-1}D^+(f_l,y)[G_{\rm r}(f_k,y)D(f_{k+1},y)\nonumber\\
&&~~~~~~~~~~~~~~~~~~~~~~~~-D(f_k,y)G_{\rm r}(f_{k+1},y)]\prod_{l=k+2}^{n}D^+(f_l,y)\, dy \, .
\end{eqnarray}
For all $y\in \M$ one of the following cases
holds true:

(I) $\bar V_y^+\cup \bar V_y^-\cup\{y\}$ neither intersects $\supp f_k$ nor $\supp f_{k+1}$: In this case $D(f_k,y)=D(f_{k+1},y)=0$ and the
term in the brackets $[\ldots]$ on the r.h.s. of (\ref{4.2eqa}) vanishes.

(II) $y\in\supp f_k$: From $\supp f_k\perp \supp f_{k+1}$ it follows that $D(f_{k+1},y)=G_{\rm r}(f_{k+1},y)=0$ and the bracket $[\ldots]$ in (\ref{4.2eqa})
again vanishes.

(III) $\bar V_y^-$ and $\bar V_y^+$ intersect $\supp f_k$: As in (II) one concludes $y\perp \supp f_{k+1}$ and $[\ldots]=0$.

(IV) Only $\bar V_y^-$ intersects $\supp f_k$ and $y\not\in\supp f_k$: Then $\supp f_k\perp\supp f_{k+1}$ $\Rightarrow$ $\supp f_{k+1}\cap (V^+_y\cup\{y\})=\emptyset$ hence
$G_{\rm r}(f_k,y)=G_{\rm r}(f_{k+1},y)=0$ and again $[\ldots]=0$.

(V) Only $\bar V_y^+$ intersects $\supp f_k$ and $y\not \in \supp f_k$: One replaces the expressions $G_{\rm r}$ in the bracket with $G_{\rm a}$ via adding a term
$[D(f_{k+1},y)D(f_{k},y)-D(f_{k},y)\linebreak \times D(f_{k+1},y)]=0$, cf. (\ref{2.2eqa}), and then concludes as in (IV) that $G_{\rm a}(f_k,y)=G_{\rm a}(f_{k+1},y)=0$ $\Rightarrow$ $[\ldots]=0$.  

Hence the bracket $[\ldots]$ in (\ref{4.2eqa}) vanishes for all $y\in \M$. 
This establishes Theorem \ref{4.1theo}.

Without pretending to be mathematically rigorous, we want to indicate, why in general positivity of the Wightman functions can not be expected. We give an argument similar to the Jost-Schroer theorem \cite{SW} in the Minkowski case.

 Let us assume for a moment that positivity holds.
One then gets the vacuum representation of the algebra of local fields through the well-known Wightman reconstruction theorem \cite{SW}. Let us furthermore assume that
the vacuum is separating for the local fields -- rather general sufficient conditions which imply this ("Reeh-Schlieder property") can be found in \cite{SVW}.
It then follows from (\ref{2.4eqa}) that $\langle j(f)\Psi_0,j(f)\Psi_0\rangle=0$,  $\forall f\in {\cal D}(\M,\C)$,
hence $j(f)\Psi_0=0$ and $j(f)=0$. From the Yang-Feldman equations (\ref{3.1eqa}) one then gets $\phi=\phi^{\rm in}$ in contradiction with (\ref{3.6eqa}). 

\section{Scattering behavior}
 \label{6sect}
 In this section we determine the (non-trivial) scattering behavior
 of the theory and the properties and representation of the outgoing field. Solving the Yang-Feldman
 equations (\ref{3.1eqa}) for the "out"-field, we obtain for $a_1,\ldots,a_n=$in/loc/out:
 \begin{eqnarray}
 \label{5.1eqa}
 &&\left\langle\Psi_0,\phi^{a_1}(f_1)\cdots\phi^{a_n}(f_n)\Psi_0\right\rangle^T\nonumber\\
 &=&b_n\sum_{k=1:a_k={\rm loc}}^n\int_\M\prod_{l=1}^{k-1}D^-(f_l,y)G_{\rm r}(f_k,y)\prod_{l=k+1}^nD^+(f_l,y)\, dy\nonumber\\
 &+&b_n\sum_{k=1:a_k={\rm out}}^n\int_\M\prod_{l=1}^{k-1}D^-(f_l,y)D(f_k,y)\prod_{l=k+1}^nD^+(f_l,y)\, dy\, .
 \end{eqnarray} 
Here all integrals, cf. Condition \ref{3.1cond}. Just as in section \ref{5sect} one can show that
(\ref{5.1eqa}) is Hermitean and invariant under ${\cal G}^\uparrow(\M,g)$. As we shall show below, locality hold for each of the fields -- "in", "loc" and "out"-field --
separately, but of course not jointly. 

The equation (\ref{5.1eqa}) in particular determines the (truncated)
scattering-(S-)matrix elements
\begin{eqnarray}
\label{5.2eqa}
&&\left\langle\phi^{\rm in}(\bar f_k)\cdots\phi^{\rm in}(\bar f_1)\Psi_0,\phi^{\rm out}(f_{k+1})\cdots\phi^{\rm out}(f_n)\Psi_0\right\rangle^T\nonumber\\
 &=&b_n{i\over 2}\left\{\int_\M\prod_{l=1}^{k}D^-(f_l,y)\prod_{l=k+1}^nD^+(f_l,y)\, dy
 -\int_\M\prod_{l=1}^{n}D^-(f_l,y) dy\right\} .
\end{eqnarray}
Here we used Hermiticity of the fields (see below and Appendix \ref{Aapp}) and $D=(-i/2)(D^+-D^-)$ to simplify (\ref{5.1eqa}) in this special case using "telescope" cancellations. 

Next we want to discuss questions concerning the algebraic properties of "in"- and especially "out"-fields. 
(\ref{2.4eqa}), (\ref{3.2eqa}) and (\ref{5.1eqa}) give the collection of mixed non-truncated vacuum expectation values of "in"-, "loc"- and "out"-fields, called the form factor functional \cite{AG}. One can now use the standard GNS-construction on inner product spaces to 
obtain a representation of the algebra generated from "in"- "loc"- and "out"-fields on some non-degenerate inner product space $({\cal V},\langle.,.\rangle)$ with a specific GNS-"vacuum" vector $\Psi_0\in{\cal V}$.  This
gives a precise mathematical meaning to the fields $\phi^{\rm in}(f),\phi(f),\phi^{\rm out}(f)$ as operator valued distributions on ${\cal V}$ (the domain of definition is the entire space) and justifies writing the left hand side of (\ref{5.1eqa}) as a truncated
 vacuum expectation value w.r.t. the inner product $\langle.,.\rangle$. For the details we refer to the Appendix \ref{Aapp}.

That the incoming field $\phi^{\rm in}$ fulfills the Klein-Gordon equation and the CCR was part of our {\em ansatz}, cf. (\ref{3.4eqa}). If the scattering defined in our model is reasonable, the same properties 
should also hold for the "out"-field:

\begin{theorem}
\label{5.1theo}
The outgoing field $\phi^{\rm out}$ fulfills the Klein--Gordon equation and the CCR on the entire state space ${\cal V}$.
\end{theorem}

It is rather easy to verify the Klein-Gordon equations for $\phi^{\rm out}$: If we apply the Klein-Gordon operator $\Box+m^2$ in (\ref{5.1eqa}) for $a_k=$out to $f_k$, then the argument
$(\Box+m^2)f_k$ in the first sum on the right hand side stands in one of the functions $D^+$ or $D^-$ and in the second sum on the
right hand side in one of the functions $D^+$, $D^-$ or $D$. As all these fundamental functions fulfill the Klein-Gordon equation, cf. (\ref{2.2aeqa}) and (\ref{2.3eqa}) the assertion follows for the
truncated $n$-point functions, $n\geq 3$. But it also holds for the truncated two point function $(c^2/m^2)D^+$ as it fulfills this equation in both arguments. If we now go over from truncated
to non-truncated vacuum expectation values, the test function $f_k$ occurs in exactly one truncated $n$-point function and hence the vacuum expectation value vanishes if the Klein-Gordon operator is applied.
This proves $\langle\Psi,\phi^{\rm out}((\Box+m^2)f)\Phi\rangle=0$ $\forall \Psi,\Phi\in{\cal  V}$. As the inner product on ${\cal V}$ is non-degenerate, $\phi^{\rm out}((\Box+m^2)f)\Phi=0$ $\forall \Phi\in{\cal V}$ follows.
By definition, this means $\Box \phi^{\rm out}+m^2\phi^{\rm out}=0$.

Next we prove the CCR. The following lemma, connecting CCR and truncation on a general level, is needed:
\begin{lemma}
\label{5.1lem}
For the CCR for $\phi^{\rm out}$ to hold it is necessary and sufficient that for $n\geq 3$, $k=1,\ldots,n-1$, $a_1,\ldots,a_{k-1},a_{k+2},\ldots,a_n\in\{ {\rm in,loc,out} \}$, $f_1,\ldots,f_n\in{\cal D}(\M,\C)$ arbitrary
\begin{equation}
\label{5.3eqa}
\left\langle  \Psi_0,\phi^{a_1}(f_1)\cdots[\phi^{\rm out}(f_k),\phi^{\rm out}(f_{k+1})]\cdots\phi^{a_n}(f_n)\Psi_0\right\rangle^T=0.
\end{equation}
\end{lemma}
The proof of Lemma \ref{5.1lem} can be found in Appendix \ref{Bapp}. We thus have to verify the sufficient condition (\ref{5.3eqa}). If we calculate the left hand side of this equation using (\ref{5.1eqa})
we obtain (up to a sign) the same expression as on the right hand side of (\ref{4.3eqa}) with $G_{\rm r}$ replaced by $D$. But then it follows from the expression in the brackets $[\cdots]$ on the right hand side of
(\ref{4.3eqa}) that after this replacement the expression vanishes identically for arbitrary test functions $f_k,f_{k+1}$ (with not necessarily space-like separated support).  This proves Theorem \ref{5.1theo}.

Having proven the main features of the free field for the "out"-field, it remains to investigate the representation of the CCR given by the "out"-field. The truncated two-point function for $\phi^{\rm in}$ and $\phi^{\rm out}$ coincides,
cf. (\ref{3.2eqa}).  For the truncated $n$-point functions, $n\geq3$, we however find similarly as in (\ref{5.2eqa}) through "telescope" cancellations for real-valued test-functions $f_1,\ldots,f_n\in{\cal D}(\M,\R)$
\begin{eqnarray}
\label{5.4eqa}
&&\left\langle\Psi_0,\phi^{\rm out}(f_1)\cdots\phi^{\rm out}(f_n)\Psi_0\right\rangle^T\nonumber\\
&&~~~~~~~~~~~~~~~~~~~~~~=b_n{i\over 2}\left[\int_\M\prod_{l=1}^nD^+(f_l,y) \,dy-\int_\M\prod_{l=1}^nD^-(f_l,y) \, dy\right]\nonumber\\
&&~~~~~~~~~~~~~~~~~~~~~~=b_n\,{\rm Im}\left[\int_\M\prod_{l=1}^nD^-(f_l,y)\, dy\right].
\end{eqnarray}
Hence, the representations of the CCR given by the field $\phi^{\rm in}$ and the one given by $\phi^{\rm out}$ are unitary equivalent\footnote{In the restrictive sense that there exists a linear isometry (w.r.t. $\langle.,.\rangle$) $V$ from the "in"-space
generated by application of the "in"-fields to the vacuum to the related "out"-space s.t. $V\Psi_0=\Psi_0$ and $V\phi^{\rm in}V^{-1}=\phi^{\rm out}$, cf. \cite{SW}. In \cite{Wa} the notion is used in the larger sense that not necessarily $V\Psi_0=\Psi_0$.} only if the right hand side of
(\ref{5.4eqa}) vanishes. Sufficient conditions for this will be given in the next section -- they apply to the case of stationary space-times. Hence, a non-vanishing of the
right hand side of (\ref{5.4eqa}) can be seen as a consequence of the interaction of a non-stationary space-time (a time dependent classical gravitation) with the quantum scattering
due to non-vanishing truncated $n$-point functions, $n\geq 3$, leading to a non-Fock and not quasi-free representation for the "out"-field.

\section{The case of stationary space-times}
\label{7sect}

In this section we consider the special case
of $\M$ being a stationary space-time, i.e. the metric $g$ in the coordinates $(t,\vec x)\in\R\times\Sigma$ does not depend on $t$. 
In the described situation time translations form a one parameter group of symmetries and the fundamental functions,
e.g. $D^+$ only depend on the differences of time arguments: $D^+(x,y)=D^+(x^0-y^0,\vec x,\vec y)$. We then define the Fourier transform\footnote{Throughout
this section we assume that all fundamental functions and all vacuum expectation values (\ref{5.1eqa}) are tempered distributions in the time arguments s.t. Fourier transforms 
are well defined as Fourier transforms of tempered distributions. This property of course has to be verified for given $(\M,g)$ and $D^+$.}  
\begin{equation}
\label{6.1eqa}
\hat D^+(f,h_1,h_2)=D^+(\hat f,h_1,h_2),~~~~\hat f(E)=(2\pi)^{-1/2}\int_{\R}e^{-itE} f(t)\, dt,
\end{equation}
$f\in{\cal S}(\R),\, h_1,h_2\in{\cal D}(\Sigma,\C)$ with ${\cal S}(\R)$ the space of complex-valued Schwartz functions on $\R$ and ${\cal D}(\Sigma,\C)=C_0^\infty(\Sigma,\C)$. The Fourier transforms $\hat D^-,\hat G_{\rm r/a}$ 
of the remaining fundamental functions are defined analogously
and the Fourier transform of vacuum expectation values (\ref{5.1eqa}) is defined by taking the Fourier transform in each time argument $t_l=x_l^0$, $l=1,\ldots,n$.
We sometimes suppress the test functions $h_1,h_2\in {\cal D}(\Sigma,\C)$ if they do not matter in a specific argument.

Let $\{{\sf U}(t)\}_{t\in\R}$ be the unitary (w.r.t. the indefinite inner product $\langle.,.\rangle$) representation of the one parameter group
of time translations, cf. Theorem \ref{4.1theo} and Apendix \ref{Aapp}. In such a situation spectral conditions can be formulated as follows:

\begin{condition}
\label{6.1cond}
(i) $D^+$ fulfills the spectral condition with spectral gap $\epsilon>0$ if the Fourier
 transform in the time variable, $\hat D^+(E)$, vanishes for $E< \epsilon$, i.e. $\supp \hat D\cap (-\infty,\epsilon)=0$.

\noindent (ii) The indefinite metric QFT over $(\M,g)$ constructed in Section \ref{6sect} fulfills the spectral condition if $\forall \Phi,\Psi\in{\cal V}$
\begin{equation}
\label{6.2eqa}
\int_{\R} \left\langle \Psi,{\sf U}(t)\Phi\right\rangle f(-t)\, dt=0 \, , ~~~\forall f\in{\cal S}(\R),~\supp \hat f\cap (-\infty,0)=0.
\end{equation}
\end{condition}
Condition \ref{6.1cond} (i) fixes $D^+$ uniquely, cf. \cite[Chapter 3.3]{Wa}.

The spectral condition (ii) means that the spectrum of the
generator of time translations, i.e. the spectrum of the Hamiltonian, is bounded from below with lowest eigenvalue zero (assumed by the vacuum). The spectral condition (i) means that this is only true
for the space of "one particle states", i.e. the sub-space of ${\cal V}$ obtained by applying once the local field to the vacuum. . Thus, (i) is weaker than (ii). In our model we however also have that
(i) implies (ii):

\begin{theorem}
\label{6.1theo}
Let $(\M,g)$ be a stationary space-time and let $D^+$ fulfill the spectral condition \ref{6.1cond} (i). Then,
the spectral condition is fulfilled for the entire theory, cf. Condition \ref{6.1cond} (ii).
Furthermore, the representations of the CCR generated
by {\rm "in"}- and {\rm "out"}-fields are unitary equivalent.
\end{theorem}

We first prove the spectral property (\ref{6.2eqa}). By the same methods as in the flat case \cite{SW} one can show that the spectral condition \ref{6.1cond} (ii) 
is equivalent to the spectral condition for the vacuum expectation values
in the sense that the support of their Fourier transform in the time arguments lies
in the set $\{ (E_1,\ldots,E_n)\in\R^n:\sum_{l=1}^nE_l=0,\sum^n_{l=r}E_l\geq 0,r=2,\ldots,n\}$. Furthermore, this spectral condition
for the vacuum expectation values is equivalent to the spectral condition for the truncated vacuum expectation values, cf. \cite{Ar,BLOT}.
For the two-point function this is just Condition \ref{6.1cond} (i). We therefore only have to verify this support property for the expressions (\ref{5.1eqa}).

Let us Fourier transform any term in the first sum on the right hand side of (\ref{5.1eqa}). Surpressing $h_l\in{\cal D}(\Sigma,\C),\vec y\in\Sigma$ arguments and the $\int_\Sigma\cdots d\vec y$ integration, the result is up to a constant $b_n/2\pi$
\begin{equation}
\label{6.3eqa}
\prod_{l=1}^{k-1} \hat D^-(E_l)\, \hat G_{\rm r}(E_k)\prod_{l=k+1}^n\hat D^+(E_l)\, \delta(\sum_{l=1}^nE_l).
\end{equation}
Note that the product of distributions in (\ref{6.3eqa}) is well defined by Condition \ref{3.1cond}.
For $r=k+1,\ldots,n$, $\sum_{l=r}^nE_l>0$ on the support of the above expression as each $E_l>0$, $l=r+1,\ldots,n$. Let thus $r\leq k$. We note that
$\hat D^-(f,h_1,h_2)=\overline{\hat D^+( \bar f_\theta,\bar h_1,\bar h_2)}$, $f_\theta(E)=f(-E)$, $f\in{\cal S}(\R)$, $h_1,h_2\in{\cal D}(\Sigma,\C)$, and thus $\supp\hat D^-\cap (-\infty,-\epsilon)=\emptyset$. In the support of the distribution (\ref{6.3eqa}) we thus have $E_l<0$, $l=1,\ldots,k-1$ and consequently
$\sum_{l=r}^nE_l=-\sum_{l=1}^{r-1}E_l>0$. 

 The terms in the second sum on the r.h.s. of (\ref{5.1eqa}) can be treated analogously, as in the above argument we did not need any properties of $\hat G_{\rm r}$. This establishes the spectral condition.

In order to prove unitary equivalence of the CCR representations for "in"- and "out"-fields, we have to show that the right hand side of (\ref{5.4eqa}) vanishes. 
Taking the Fourier transform in the time arguments of the term in the brackets in (\ref{5.4eqa}) yields
\begin{equation}
\label{6.xxeqa}
\prod_{l=1}^n \hat D^-(E_l)\, \delta(\sum_{l=1}^nE_l)=0\, ,
\end{equation}
as the delta function and the product have disjoint support. Here again we suppressed a factor $1/2\pi$, arguments $h_l\in{\cal D}(\Sigma,\C),\vec y$ and $d\vec y$-integration over $\Sigma$. This proves the theorem.

By the same argument we also get that in the stationary case only the first term
in the curly brackets on the right hand side of (\ref{5.2eqa}) gives a non vanishing contribution. 

In particular, this applies
to the case of Minkowski space-time, where we recover the same scattering behavior as in \cite{AG}. The first term on the right hand side of (\ref{5.2eqa})
in energy-momentum space then just gives the on-shell and energy-momentum conservation term (up to a constant) describing somehow the "simplest possible"
non-trivial scattering behavior. 

Another immediate consequence of Theorem \ref{6.1theo} follows from the fact that the distribution $D^+$ is positive, hence "in"- and "out" fields create positive Fock representations: 
\begin{corollary}
\label{6.1cor}
Under the conditions of Theorem \ref{6.1theo} one gets that
the restriction  of the inner product $\langle.,.\rangle$ to the spaces ${\cal V}^{\rm in/out}\subseteq{\cal V}$ generated by repeated application of the asymptotic fields $\phi^{\rm in/out}$ to the vacuum 
is positive semi-definite.
\end{corollary}

\section{Verifications for de Sitter space-time}
 \label{8sect}
In this section we want to consider de Sitter spaces as a concrete example of 
 curved space-times. The choice of the Sitter spaces is particularly interesting
 in the respect that Condition 3.1 sensitively depends on the dimension, see Theorem 
 \ref{condsitter}. 
 Only for dimensions $d\geq 6$ Condition 3.1 can be verified for all orders of 
 the Wightman functions. In lower dimensions it may well depend on the order, as for 
 example, for dimension 4  the third order does not exist. This is an infrared problem 
 which has also been observed by Tagirov \cite{Tag} in the context of $:\varphi^3:$ theory.
 In the present case this could be repaired by simply choosing $b_3=0$.
 
 Note that the Sitter spaces have
 spheres as Cauchy surfaces and this compactness at space-like distances hinders
 dispersion. On the other hand the volume of de Sitter spaces increases rather fast
 when moving along the time-like direction, which may facilitate dispersion.
 A more careful treatment given below than shows that these effects really seem 
 to be responsible for whether Condition 3.1 holds or does not hold.
   
 The choice of de Sitter spaces is also convenient for the discussion in as far as there is 
 a preferred vacuum, the so called Euclidean or Bunch-Davies vacuum \cite{BuDa}. It is the 
 distinguished one, which is selected from other choices by the demand of covariance
 and the Hadamard condition, see e.g. \cite{All,BrMo}. Note that de Sitter spaces are maximally
 symmetric, i.e. the dimension of the symmetry group is maximal, hence covariance is 
 a natural axiom to be imposed on the Wightman functions. The Hadamard condition then 
 selects the particular two-point function which has the same singular behavior at
 light-like distances as the two-point function of the free field in Minkowski space.\\
 Below we shall prove the following 
\begin{theorem}\label{condsitter}
Given the $d-$dimensional de Sitter space-time $\Bbb{X}^d$ and the Bunch-Davies
vacuum $D^+$ on it, Condition 3.1 holds if $(dn-2n-2d)/2 >-1$, $n\geq 3$.
\end{theorem} 
In order to investigate Condition 3.1, we shall use essentially the results given 
 in \cite{ChTa,Tag}, where an orthonormal mode expansion is used for the representation of 
 the two-point function. Let us recall some basic features.
 Given the $(d+1)-$dimensional Minkowski space $(\Bbb{M}^{d+1}, g)$ with metric 
 $g=g_{\mu\nu}dx^\mu dx^\nu$, where $g_{\mu\nu}={\rm diag}(+1,-1, \ldots,-1)$. 
 The $d$-dimensional de Sitter space $(\Bbb{X}^d, \overline{g})$ is defined as
 $$\Bbb{X}^d := \left\{x \in \Bbb{M}^{d+1}|\; g_{\mu\nu}x^\mu x^\nu =
 (x^0)^2-(x^1)^2-\cdots-(x^d)^2=-r^2\right\}, \quad r >0,
 $$ 
 equipped with the metric $\overline{g}=\iota^\ast g$, where $\iota^\ast$ 
 denotes the pull-back
 with respect to the imbedding $\iota:\Bbb{X}^d \rightarrow \Bbb{M}^{d+1}.$
 The eigenmodes are calculated in "global" coordinates, which in fact are given
 by the diffeomorphisms
 \begin{equation}
 \kappa_d: \left\{
 \begin{array}{ccl}(-\frac{\pi}{2},\frac{\pi}{2})\times \Bbb{S}^{d-1} &
 \longrightarrow & \Bbb{X}^d \\
 (\tau,\alpha)&\longmapsto & (r \tan \tau, r \alpha /\cos \tau).
 \end{array} \right.
 \end{equation}
 The pull-back of the volume form $dV^d$ on $\Bbb{X}^d$ is then given by \\
 $\kappa_d^\ast \, dV^d = \cos^{-d}\tau d\tau d\Omega^{d-1}$, where $d\Omega^{d-1}$ is the 
 volume form on the sphere $\Bbb{S}^{d-1}.$ 
 In these coordinates the Klein-Gordon equation $(\Box +m^2)\varphi = 0$ reads
 \begin{equation} \label{Kleinco}
 \left(\cos^d\tau\partial_\tau \cos^{2-d}\tau\partial_\tau  -
 \cos^2\tau \Delta_{\Bbb{S}_{d-1}}  + \frak{m}^2\right)\varphi =0, \; \frak{m}=mr.
 \end{equation}
 Solutions to (\ref{Kleinco}) can be found by separation of variables. Setting $\varphi(\tau,\alpha)=
 T(\tau)\Xi(\alpha)$ gives
 \begin{equation}\label{harmonic}
 (\Delta_{\Bbb{S}^{d-1}}+\kappa^2)\Xi =0,
 \end{equation}
 and
 \begin{equation}\label{taupart}
\left(\cos^d\tau \partial_\tau \cos^{2-d}\tau\partial_\tau -
 \kappa^2 \cos^2\tau  +\frak{m}^2\right)T =0,
 \end{equation}
 where $\kappa^2$ is the separation constant.
Note that $L^2(\Bbb{S}^{d-1})\simeq \bigoplus_{s=0}^\infty {\cal H}^s(\Bbb{S}^{d-1}),$
where the subspaces ${\cal H}^s(\Bbb{S}^{d-1})$ are spanned by the spherical harmonics
$(\Xi^s_l)_{0\leq l \leq h(s,d)}$ of degree $s$, $h(s,d)={\rm dim}{\cal H}^s(\Bbb{S}^{d-1})$,
which at the same time are eigensolutions of (\ref{harmonic}) with
eigenvalues $-\kappa^2 = -s(s+d-2).$ 
Moreover, the quasi-regular representation $Q:{\rm SO}(d)\rightarrow L^2(\Bbb{S}^{d-1}),$
given by $Q(k)(f)(\alpha)=f(k^{-1}\alpha),$ splits into a direct sum of unitary irreducible representations
$Q^s$ on ${\cal H}^s(\Bbb{S}^{d-1})$, such that $Q = \bigoplus_{s=0}^\infty Q^s$. If the $q^s_{uv}$
mean the matrix elements of $Q^s$ with respect to the bases $(\Xi^s_l)_{0\leq l \leq h(s,d)}$,
then the following relation holds \cite[p. 470]{Vil}
\begin{equation}\label{relation}
\Xi^s_l(\alpha) = ({\rm dim}{{\cal H}^s(\Bbb{S}^{d-1})})^{1/2}\overline{q^s_{l0}}(k),
\end{equation}  
where $\alpha=ke_n$, $k \in {\rm SO}(d)$, $e_n=(0,0,\ldots,1)$ being the invariant vector with respect to the 
subgroup ${\rm SO}(d-1)$, so that $\Bbb{S}^{d-1}\simeq {\rm SO}(d)/{\rm SO}(d-1)$.
Pairs of linearly independent solutions to
(\ref{taupart}) are given by \cite{ChTa}
\begin{equation}
T^\pm_p(\tau)=\cos^{(d-2)/2}\tau u^\pm_p(\tau),\; p=s+(d-2)/2,
\end{equation}
where 
\begin{equation}
\begin{split}
 u^\pm_p(\tau)=&(p!)^{-1}\sqrt{\Gamma(p+\mu)\Gamma(p-\mu+1)}
e^{\pm i p \tau} F(\mu, 1-\mu, p+1; e^{\pm ip\tau}/(2 \cos\tau)),\\
& \mu =1/2(1-\sqrt{(d-1)^2 -4\frak{m}^2}),
\end{split}
\end{equation}
$F$ being the hypergeometric function.
We thus get the following system of solutions
\begin{equation}
\varphi^\pm_{pl}(x)=T^\pm_p(\tau)\Xi^s_l(\alpha), \quad 0 \leq p < \infty,
\; 0 \leq l \leq h(s,d).
\end{equation} 
In terms of these solutions the two-point function is expressed as
\begin{equation}\label{two-point}
\begin{split}
& D^+(x_1,x_2)= \sum_{s=0}^\infty\sum_{0\leq l \leq h(s,d)}\varphi^+_{pl}(x_1)
\varphi^-_{pl}(x_2)=\\
& \cos^{(d-2)/2}\tau_1\cos^{(d-2)/2}\tau_2
\sum_{s=0}^\infty u^+_p(\tau_1)u^-_p(\tau_2)A(s,d)
C^{1/2(d-1)}_s(\cos(\alpha_1,\alpha_2)),
\end{split}
\end{equation}
with 
\begin{equation}
\sum_{0 \leq l \leq h(s,d)} \Xi^s_l(\alpha_1)
\overline{\Xi_l^s}(\alpha_2) = A(s,d)C_s^{1/2(d-2)}(\cos(\alpha_1,\alpha_2)),
\end{equation}
where $C_s^{1/2(d-2)}$ is a Gegenbauer polynomial, $A(s,d)=(2s+(d-2))\Gamma(1+1/2(d-2))/(
2\pi^{1+1/2(d-2)}(d-2))$, and
$(\alpha_1,\alpha_2)$ denotes the angle between $\alpha_1$ and $\alpha_2$.
The convergence of the series (\ref{two-point}) has to be understood
in the weak topology of ${\cal D}(\Bbb{X}^d \times\Bbb{X}^d,\Bbb{C})'$. We will see that 
$D^+(x_1,x_2)$ is even an element in the subspace 
$L({\cal D}(\Bbb{X}^d,\Bbb{C});C_b(\Bbb{X}^d,\Bbb{C}))
$, the space of continuous linear mappings from ${\cal D}(\Bbb{X}^d,\Bbb{C})$ 
to the Banach space of complex-valued bounded continuous functions on $\Bbb{X}^d$, equipped with the 
supremum norm.
This means that applying (or 
smearing with) a test function the series will not only converge to a 
function in $C_b(\Bbb{X}^d,\Bbb{C})$, but the result will depend continuously on
the chosen test functions. At the same time Condition 3.1 will be verified. 
So let us smear (\ref{two-point}) with an $f\in {\cal D}(\Bbb{X^d},\Bbb{C})$ in one 
argument, say $x_1$.
In order to have control 
on the summation we need the following asymptotic formulas \cite[Ch. 2.2.2 and 1.18]{Erd}
\begin{equation}\label{asym1}
F(\mu,1-\mu,p+1,e^{\pm ip\tau}/(2\cos\tau))=1 + O(1/p),
\end{equation}
and
\begin{equation}\label{asym2}
(p!)^{-2} \Gamma(p+\mu)\Gamma(p-\mu+1)=1/p + O(1/p^2).
\end{equation}
(\ref{asym1}) holds on every compact interval $\left[a,b\right]\subset(
-\frac{\pi}{2},\frac{\pi}{2})$.
Using (\ref{asym1}) and (\ref{asym2}) in (\ref{two-point}), we need to investigate the 
series

\begin{equation}\label{subserieso}
\begin{split}
 & D_{[a,b]}(\tau_2,\alpha_2)=1_{[a,b]}(\tau_2) \cos^{(d-2)/2}\tau_2\times \\
 & \sum_{s=0}^\infty\sum_{0 \leq l \leq h(s,d)}\int_{(-\frac{\pi}{2},\frac{\pi}{2})
 \times \Bbb{S}^{d-1}}
(1/p + O(1/p^2))\Xi^s_l(\alpha_1)
\overline{\Xi_l^s}(\alpha_2)\times \\
 & e^{i (d-2)\tau_1/2}e^{i s \tau_1} e^{-i s \tau_2}e^{-i(d-2)\tau_2/2} 
f(\tau_1,\alpha_1)\cos^{(-d-2)/2}\tau_1 d \tau_1 d\Omega^{d-1}(\alpha_1).
\end{split}
\end{equation}
If we consider the similar expression without the terms 
$1/p+O(1/p^2)$,
\begin{equation}\label{subseries}
\begin{split}
 & \tilde{D}(\tau_2,\alpha_2)= \cos^{(d-2)/2}\tau_2\times \\
 & \sum_{s=0}^\infty\sum_{0 \leq l \leq h(s,d)}\int_{(-\frac{\pi}{2},\frac{\pi}{2})
 \times \Bbb{S}^{d-1}}
\Xi^s_l(\alpha_1)
\overline{\Xi_l^s}(\alpha_2)\times \\
 & e^{i (d-2)\tau_1/2}e^{i s \tau_1} e^{-i s \tau_2}e^{-i(d-2)\tau_2/2} 
f(\tau_1,\alpha_1)\cos^{(-d-2)/2}\tau_1 d \tau_1 d\Omega^{d-1}(\alpha_1).
\end{split}
\end{equation}  
then up to the factors $e^{i (d-2)\tau_1/2},\; 
e^{-i(d-2)\tau_2/2}$ it represents a sub-series of the harmonic
expansion of the function $\tilde{f}(\tau_1,\alpha_1):=
e^{i(d-2)\tau_1/2}f(\tau_1,\alpha_2)\cos^{(-d-2)/2}\tau_1,$\\ $\tilde{f} \in 
{\cal D}(\Bbb{S}^1\times\Bbb{S}^{d-1},\Bbb{C})$, if we identify $\Bbb{S}^1\simeq \Bbb{R}
/(2\pi\Bbb{Z}+\pi)$ and extend $\tilde{f}$ to $[-\pi,\pi]\times \Bbb{S}^{d-1}$ by setting it equal 
zero outside its support. Using the relation (\ref{relation}) we may regard 
this modified expression as a series
on the 
compact Lie group $\Bbb{S}^1\times {\rm SO(d)}$. Before proceeding we recall the 
following facts from harmonic analysis on Lie groups. 
Let a compact Lie group $K$ be given. Let $\hat{K}$ denote the equivalence classes of 
irreducible unitary representations of $K$. The representatives $U_\lambda, \,\lambda
\in \hat{K}$ are finite-dimensional with dimension denoted $d(\lambda)$.
We may write $u^\lambda_{ij}(k)$ for the matrix of $U_\lambda(k), \; k\in K,$ after having choosen some 
basis. According to the theorem of 
Peter-Weyl any  
$f\in L^2(K)$ has the following series or harmonic expansion in the $L^2$-sense 
\cite[Theorem 26.40]{HeRo},

\begin{equation} \label{peter}
f=\sum_{\lambda \in \hat{K}} \sum_{i,j=1}^{d(\lambda)}d(\lambda)(f,u^\lambda_{ij})u^\lambda_{ij},
\end{equation}
where $(f,u^\lambda_{ij})=\int_K f \overline{u}^\lambda_{ij}dk$. The integration is 
performed with respect to the Haar measure $dk$ and the bar means complex conjugation.
(\ref{peter}) remains true when switching to complex conjugates.
For functions $f\in C^\infty(K,\C)$ this statement can be sharpened a lot \cite[Theorem 1]{Sug}. 
Let $w$ denote the dimension of the 
maximal toral subgroup of $K$ and let ${\rm dim}\,K = w + 2\gamma$. For
$f \in C^\infty(K,\C)$ the series (\ref{peter}) converges absolutely and uniformly
with the estimate
\begin{equation}\label{est}
\sum_{\lambda\in \hat{K}} \sum_{i,j=1}^{d(\lambda)}d(\lambda)|(f,u^\lambda_{ij})|
\left\|u^\lambda_{ij}\right\|_\infty
\leq \left\|\Delta_K^l f\right\|_2(N^2\sum_{\lambda \in \hat{K}}
|\lambda|^{2\gamma-4l})^{1/2},\;
\forall 2l>w/2+\gamma,
\end{equation}
where $\left\|\cdot\right\|_\infty$ denotes the supremum norm, $N$ is a constant, $\Delta_K$ is the Laplacian on $K$ and $\left\|\cdot\right\|_2$ denotes 
the $L^2(K)$-norm. \\
Now the series of the absolute values of (\ref{subseries}) can be identified with 
a sub-series of the series of absolute values of the harmonic expansion of $\tilde{f}$ 
on the
Lie group $K = \Bbb{S}^1\times {\rm SO}(d)$, which makes (\ref{subseries}) converge 
absolutely and
uniformly with estimate (\ref{est}). From the absolute uniform convergence 
of (\ref{subseries}) we obtain absolute uniform convergence of (\ref{subserieso}) together 
with the bound
\begin{equation}
|D_{[a,b]}(\tau_2,\alpha_2)| \leq \cos^{(d-2)/2}\tau_2C_f(\tau_2,\alpha_2),
\end{equation}
where $C_f$ is a continuous function on $\Bbb{S}^1\times {\rm SO}(d-1)$.
But for every $x_2=(\tau_2,\alpha_2)$ we have $\tau_2 \in [a,b]$ for appropriate
$-\frac{\pi}{2}<a<b<\frac{\pi}{2}$, so we get

\begin{equation}\label{estD}
|D^+(f,x_2)|\leq \cos^{(d-2)/2}\tau_2C_f(\tau_2,\alpha_2),
\;\forall x_2 \in \Bbb{X}^d. 
\end{equation}
On the other hand, if a sequence $(f_l)_{l \geq 0}$ converges to zero in the topology of 
$C^\infty(\Bbb{S}^1\times {\rm SO}(d),\C)$, then one easily establishes the convergence 
to zero
of the corresponding sequence $(\tilde{f}_l)_{l \geq 0}$ in the same topology. 
The topology of $C^\infty(\Bbb{S}^1\times {\rm SO}(d),\C)$ is equivalent to the topology
generated by the seminorms $p_j(f)=\left\|\Delta_{\Bbb{S}^1\times {\rm SO}(d)}^j(f)
\right\|_\infty$, hence we may deduce by (\ref{est}) that 

\begin{equation}\label{limD}
\lim_{l \rightarrow \infty}\left\|C_{f_l}\right\|_\infty=0
\end{equation}
Using (\ref{estD}) we may conclude that the expression

\begin{equation}
\int_{(-\frac{\pi}{2},\frac{\pi}{2}) \times \Bbb{S}^{d-1}}
|D^+(f,(\tau_2,\alpha_2))|^n \cos^{-d}\tau_2 d\tau_2d\Omega(\alpha_2)
\end{equation}
exists, if $(dn-2n-2d)/2\geq -1.$ Due to (\ref{limD}) also Condition \ref{3.1cond} (ii) is fulfilled with $F(x)=\sup_{n\in\N} \|C_{f_n}\|_\infty \cos^{(d-2)/2}\tau_2$. 
This proves Theorem \ref{condsitter}.

\appendix

 \section{The GNS-construction on an inner product space}
\label{Aapp}
Let $\underline{\cal D}^{ \rm ext.}$, the extended Borchers algebra, be the free tensor algebra generated by ${\cal D}^{\rm ext.}={\cal D}(\M,\C^3)$, i.e.
\begin{equation}
\label{A.1eqa}
\underline{\cal D}^{\rm ext.}=\bigoplus_{n=0}^\infty \left({\cal D}^{{\rm ext.}}\right)^{\otimes n},~~\left({\cal D}^{ \rm ext.}\right)^{\otimes 0}=\C.
\end{equation}
The addition on $\underline{\cal D}^{\rm ext.}$ is component wise and the multiplication is given by the tensor product. The involution on $\underline{\cal D}^{ \rm ext.}$ is given by
the operation $(f_1\otimes\cdots\otimes f_n)^*=\bar f_n\otimes\cdots\otimes \bar f_1$, $f_l\in {\cal D}^{\rm ext.}$, where the bar stands for complex conjugation. As we want to use this unital, involutive 
algebra to represent "in"-, "loc"- and "out"-fields, the three components of ${\cal D}^{\rm ext.}={\cal D}(\M,\C^3)$ are labeled "in" for the first component, "loc" for the second and "out" for the third. Then our
collection of mixed vacuum expectation values obtained from (\ref{2.4eqa}), (\ref{3.2eqa}) and (\ref{5.1eqa}) generates an Hermitean functional (called the form factor functional $\underline{F}$) on the extended Borchers algebra $\underline{\cal D}^{ \rm ext.}$ through
\begin{equation}
\label{A.2eqa}
\underline{F}(\underline{f})=f_0\langle\Psi_0,\Psi_0\rangle+\sum_{n=1}^\infty\sum_{a_1,\ldots,a_n\in\{\rm {in,loc,out}\}}\left\langle\Psi_0,\phi^{a_1}(f^{a_1}_{n,1})\cdots\phi^{a_n}(f^{a_n}_{n,n})\Psi_0\right\rangle,
\end{equation}
where $\underline{f}\in\underline{\cal D}^{\rm ext.}$, $\underline{f}=(f_0,\ldots,f_k,0,\ldots)$, $f_0\in\C$, $f_n=f_{1,n}\otimes\cdots \otimes f_{n,n}$, $f_{l,n}\in {\cal D}^{\rm ext.}$, and we use the normalization $\langle\Psi_0,\Psi_0\rangle=1$.

Let ${\cal L}={\cal L}(\underline{F})=\{ \underline{f}\in\underline{\cal D}^{\rm ext.}:\underline{F}(\underline{g}\otimes\underline{f})=0\, \forall \underline{g}\in\underline{\cal D}^{\rm ext.}\}$. 
Then ${\cal L}$ is a left-ideal in $\underline{\cal D}^{\rm ext.}$, i.e. $\underline{f}\in{\cal L}\Rightarrow \underline{g}\otimes\underline{f}\in{\cal L}$ $\forall \underline{g}\in\underline{\cal D}^{\rm ext.}$. In particular, $\cal L$ is a (complex) vector space
and we can define the quotient vector space ${\cal V}=\underline{\cal D}^{\rm ext.}/{\cal L}$. Let $[\underline{f}]=\underline{f}+{\cal L}$ denote the rest class of $\underline{f}$ in $\cal V$. Then,
\begin{equation}
\label{A.3eqa}
\langle[\underline{f}],[\underline{g}]\rangle=\underline{F}(\underline{f}^*\otimes\underline{g})
\end{equation}
 gives a well-defined and non-degenerate inner product $\langle.,.\rangle$ on ${\cal V}$.
  To see this, we note that by the definition of $\cal L$ the right hand side of 
  (\ref{A.3eqa}) does not depend on the choice of $\underline{g}\in[\underline{g}]$.
   By Hermiticity, $\underline{F}(\underline{f}^*\otimes \underline{g})=\overline{\underline{F}(\underline{g}^*\otimes\underline{f})}$,
    the same applies to $\underline{f}\in[\underline{f}]$. Also, if $\langle [\underline{g}],[\underline{f}]\rangle=0$ $\forall [\underline{g}]\in{\cal V}$ it follows that
	$[\underline{f}]={\cal L}$, which proves the non-degeneracy.

	The "vacuum state" $\Psi_0$, so far just a suggestive notation, is now identified with the GNS-vacuum $[(1,0,\ldots)]\in{\cal V}$. Likewise, we want to identify
	the local and asymptotic "fields" $\phi$ and $\phi^{\rm in/out}$ with operator valued distributions acting on $\cal V$.  As ${\cal L}$ is a left-ideal, we obtain a left-action
	of $\underline{\cal D}^{\rm ext.}$ on $\cal V$ through $\underline{f}\cdot[\underline{g}]=[\underline{f}\otimes\underline{g}]$ $\forall \, \underline{f}\in\underline{\cal D}^{\rm ext.}$, $[\underline{g}]\in{\cal V}$.
	Thus, every element of $\underline{\cal D}^{\rm ext.}$ can be identified with an operator on ${\cal V}$. This, in particular, applies to ${\cal D}^{\rm ext.}\subseteq \underline{\cal D}^{\rm ext.}$. Let now $\underline{f}\in {\cal D}$, i.e. $\underline{f}=(0,f^{\rm ext.},0,\ldots)$, $f^{\rm ext.}\in {\cal D}(\M,\C^3)$
	such that only the fist ("in") / second ("loc") / third ("out") component of $f^{\rm ext.}$ is different from zero and let this component be given by $f\in {\cal D}(\M,\C)$. We then define
	$\phi^{\rm in/loc/out}(f)\Psi=\underline{f}\cdot \Psi$, $\forall \Psi\in{\cal V}$. This rigorously defines $\phi^{\rm in/loc/out}(f)$ (in the text we suppress the superscript "loc" for the local field). Furthermore, by Hermiticity of $\underline{F}$, the fields $\phi^{\rm in/loc/out}$ are Hermitean w.r.t. $\langle.,.\rangle$, i.e. $\langle\Psi,\phi^{\rm in/loc/out}(f)\Phi\rangle=\langle\phi^{\rm in/loc/out}(\bar f)\Psi,\Phi\rangle$
 $\forall \Phi,\Psi\in{\cal V}$ and $f\in {\cal D}(\M,\C)$.
 
 Lastly, we want to construct a representation ${\sf U}$ of the orthochonous symmetry group ${\cal G}^\uparrow(\M,g)$. As $\underline{\cal D}^{\rm ext.}$ as a unital tensor algebra is
 generated by ${\cal D}^{\rm ext.}$, it follows that ${\cal V}$ is the linear span of vectors generated by repeated application of $\phi^{\rm in},\phi^{\rm loc}$ and $\phi^{\rm out}$ to the vacuum $\Psi_0$. To define ${\sf U}$ it is thus enough to set
 ${\sf U}(\alpha)\phi^{\rm in/loc/out}(f){\sf U}^{-1}(\alpha)=\phi^{\rm in/loc/out}(f_\alpha)$ and ${\sf U}(\alpha)\Psi_0=\Psi_0$ $\forall \alpha\in{\cal G}^\uparrow(\M,g)$. This is well defined, as the action of the symmetry group on $\underline{\cal D}^{\rm ext.}$,  
 maps ${\cal L}$ into itself, as a consequence of the invariance of $\underline{F}$ under such transformations, $\underline{F}(\underline{f})=\underline{F}( \underline{f}_\alpha)$. The invariance of the vacuum expectation values then also implies that
 ${\sf U}$ is a unitary representation ${\sf U}^*={\sf U}^{-1}$ where the adjoint is taken w.r.t. the inner product $\langle.,.\rangle$.
 
 \section{Proof of Lemma \ref{5.1lem}}
 \label{Bapp}
Let us prove that (\ref{5.3eqa}) implies the CCR. We first use the cluster expansion (\ref{2.4eqa}) for the following vacuum expectation value
\begin{eqnarray}
\label{B.1eqa}
&&\left\langle  \Psi_0,\phi^{a_1}(f_1)\cdots[\phi^{\rm out}(f_k),\phi^{\rm out}(f_{k+1})]\cdots\phi^{a_n}(f_n)\Psi_0\right\rangle\nonumber\\
&=&\sum_{I\in{\cal P}^{(n)}}\left[\prod_{\{j_1,\ldots j_l\}\in I}\left\langle  \Psi_0,\phi^{a_{j_1}}(f_{j_1})\cdots\phi^{a_{j_l}}(f_{j_l})\Psi_0\right\rangle^T\right.\nonumber\\
&&~~~~~~~~~~~~-\left. \prod_{\{j_1,\ldots j_l\}\in I}\left\langle  \Psi_0,\phi^{a_{j'_1}}(f_{j'_1})\cdots\phi^{a_{j'_l}}(f_{j'_l})\Psi_0\right\rangle^T\right].
\end{eqnarray}
Here we fixed $a_k,a_{k+1}=$out and we defined $j'_r=k+1$ if $j_r=k$, $j'_r=k$ if $j_r=k+1$ and $j_r'=j_r$ else. We can divide the partitions $I$ of $(1,\ldots,n)$ into three classes: 

1) $k$ and $k+1$ belong to different sets $A$ and $A'$ in the partition $I$. Then there exists exactly one partition $I'$ which is identical to $I$ with the exception that
$k$ and $k+1$ are exchanged\footnote{We assumed that $\langle\Psi_0,\phi^{\rm in/loc/out}(f)\Psi_0\rangle^T=0$, i.e. partitions with $A=\{k\}$ and $A'=\{k+1\}$ for which an $I'$ coincides with $I$ give a zero contribution.}. The two terms in (\ref{B.1eqa}) belonging to $I$ and $I'$ then cancel
and hence the sum over all partitions in this class gives zero.

2) $k$ and $k+1$ are in the same set $A=\{ q_1,\ldots,k,k+1,\ldots,q_r\}$ of the partition $I$ and $A$ contains more than two elements, i.e. $r\geq 3$. The summand in (\ref{B.1eqa}) belonging to such a partition is equal to
\begin{eqnarray}
\label{B.2eqa}
&&\left\langle\Psi_0,\phi^{a_{q_1}}(f_{q_1})\cdots[\phi^{\rm out}(f_k),\phi^{\rm out}(f_{k+1})]\cdots\phi^{a_{q_r}}(f_{q_r})\Psi_0\right\rangle^T\nonumber\\
&&~~~~~~~~~~~~~~~~~~~~~~~~~~~~\times\prod_{\{j_1,\ldots j_l\}\in I\setminus A}\left\langle  \Psi_0,\phi^{a_{j_1}}(f_{j_1})\cdots\phi^{a_{j_l}}(f_{j_l})\Psi_0\right\rangle^T.
\end{eqnarray}
By (\ref{5.3eqa}) the contribution from the partitions of this class also vanishes.

3) $k$ and $k+1$ are in the same set of the partition $I$ and the set contains only these two elements. The sum over all partitions in this class yields
\begin{eqnarray}
\label{B.3eqa}
&&\left\langle\Psi_0,[\phi^{\rm out}(f_k),\phi^{\rm out}(f_{k+1})]\Psi_0\right\rangle^T\nonumber\\
&&~~~~~~~~~~~~~~~~~~~~\times\sum_{I\in{\cal P}^{(n-2)}}\prod_{\{j_1,\ldots,j_l\}\in I}\left\langle\Psi_0,\phi^{a_{j'_1}}(f_{j'_1})\cdots \phi^{a_{j'_l}}(f_{j'_l})\Psi_0\right\rangle^T\nonumber\\
&&=i{b^2\over m^2}D(f_k,f_{k+1})\left\langle\Psi_0,\phi^{a_1}(f_1)\cdots\phi^{a_{k-1}}(f_{k-1})\phi^{a_{k+2}}(f_{k+2})\cdots\phi^{a_n}(f_n)\Psi_0\right\rangle. \nonumber\\
\end{eqnarray}
Here we used the notation $j_r'=j_r$ if $j_r<k$ and $j_r'=j_r+2$ if $j_r\geq k$.

From 1)--3) it follows that the left hand side of (\ref{B.1eqa}) is equal to the right hand side of (\ref{B.3eqa}). As all states in ${\cal V}$ are generated by repeated application of "in"-, "loc" and "out"-fields to the vacuum, this equality implies
$\langle\Psi,[\phi^{\rm out}(f),\phi^{\rm out}(h)]-i(b^2/m^2)D(f,h)\Phi\rangle=0$ $\forall \Phi,\Psi\in{\cal V}$. The non-degeneracy of $\langle.,.\rangle$ on ${\cal V}$ now implies
$[\phi^{\rm out}(f),\phi^{\rm out}(h)]=i(c^2/m^2)D(f,h)$. This proves the sufficiency part of the lemma.

 As we only need this part, we only sketch the necessity: If the CCR hold, then the left hand side of (\ref{B.1eqa}) is equal to the right hand side of (\ref{B.3eqa}). That this
implies (\ref{5.3eqa}) follows from the cluster expansion (\ref{2.4eqa}) by induction over $n$ taking into account that the partitions in class 1) above do not contribute to (\ref{B.1eqa}).

\
 
\noindent {\bf Acknowledgments.} Discussions with S. Albeverio, W. Junker, F. Ll\'edo and V. Moretti were very helpful to bring a minimum of
order into the materials presented here. H. G. likes to thank for
financial support via D.F.G. project "Stochastic analysis and systems with infinitely many degrees of freedom". 
H. T. gratefully acknowledges the financial support through the European TMR fellowship and the
SFB 611.


\begin{thebibliography}{11}
\bibitem{AG} S. Albeverio, H. Gottschalk: Scattering theory for quantum fields with indefinite metric, Commun. Math. Phys. {\bf 216}, 491--513 (2001).
\bibitem{AGW1} S. Albeverio, H. Gottschalk, J.-L. Wu: {Convoluted generalized white noise, Schwinger functions and their continuation to Wightman functions}, Rev. Math Phys., Vol {\bf 8}, No. {\bf 6},  763--817, (1996).
\bibitem{AGW2} S. Albeverio, H. Gottschalk, J.-L. Wu: SPDEs leading to local, relativistic vector fields with indefinite metric and non-trivial S-matrix, Proc. Trento Conf. on "Stochastic analysis", Trento 2000, eds. G. da Prato and L. Tubaro,
M. Dekker 2002.
\bibitem{All} B. Allen: Vacuum states in de Sitter space, Phys. Rev. {\bf D 32}, 3136--3149
(1985).
\bibitem{Ar} H. Araki: On the asymptotic behaviour of vacuum expectation values at large
spacelike separations. Ann. Phys. {\bf 11}, 260--274 (1960).
\bibitem{BrMo} J. Bros, U. Moschella: Two-point functions and quantum fields in de Sitter
universe, Rev. Math. Phys. {\bf 8}, 327--391 (1996).
\bibitem{BuDa} T.S. Bunch, P.C.W Davies: Quantum field theory in de Sitter space: renormalization by point-splitting
, Proc. Roy. Soc. London Ser. A {\bf 360}, 117--134 (1978).
\bibitem{BrFr} R. Brunetti, K. Fredenhagen: Microlocal analysis and interacting quantum field theories: Renormalization on physical backgrounds, Commun. Math. Phys. {\bf 208} 623--661 (2000).
\bibitem{BLOT} N. N. Bogoliubov, A. A. Logunov, A. I. Ossak, I. V. Todorov, General principles of quantum field theories, Kluwer Acad. Publ., 1990.
\bibitem{BEM} J. Bros, H. Epstein, U. Moschella, Analytic properties and thermal effects for general quantum field theory on de Sitter space-time, Commun. Math. Phys. {\bf 196}, 535--570 (1998).
\bibitem{ChTa} E. A. Chernikov, N. A. Tagirov: Quantum theory of scalar field in 
de Sitter space-time, Ann. Inst. H. Poincar\'{e}, Sect. A {\bf 9}, 109--141 (1968).
\bibitem{Di} J. Dimock, Algebras of local observables on a manifold, Commun. Math. Phys. {\bf 77}, 219--228 (1980).
\bibitem{Erd} A. Erd\'{e}lyi, editor: Higher transcendental functions, Vol. 1 and 2, 
Bateman manuscript project. New York: McGraw-Hill 1953.
\bibitem{GV} I. M. Gelfand, N. Ya. Vilenkin:
Generalized Functions, IV. Some Applications of Harmonic
Analysis. New York/London: Academic Press 1964.
\bibitem{GT} H. Gottschalk, H. Thaler, in preparation.
\bibitem{Haw} S. Hawking: Particle production by black holes. Comm. Math. Phys. {\bf 43}, 199--220 (1975).
\bibitem{HeRo} E. Hewitt, K. A. Ross: Abstract harmonic analysis II. Die 
Grundlehren der mathematischen Wissenschaften, Band 152. New-York/Berlin: Springer 
Verlag 1970.
\bibitem{HR} S. Hollands and W. Ruan, The state space of perturbative quantum field theory in curved spacetimes, Ann. H. Poincar\' e {\bf 3}, 635--657 (2002).
\bibitem{M} R. A. Minlos: Generalized random processes and their extension in measure. Translations in
Mathematical Statistics and Probability , AMS Providence {\bf 3}, 291--313 (1963).
\bibitem{MS} G. Morchio, F. Strocchi, { Infrared singularities, vacuum structure and pure phases in local quantum field theory}, Ann. Inst. H. Poincar\'e, Vol. {\bf 33}, 251--282, (1980).
\bibitem{Os} A. Ostendorf, Feynman rules for Wightman functions, Ann Inst. H. Poincar\'e {\bf 40}, 273--290 (1984).
\bibitem{Ru} D. Ruelle, {On the asymptotic condition in quantum
field theory}, Helv. Phys. Acta {\bf 35}, 147--163, (1962).
\bibitem{Ste1} O. Steinmann, Perturbation theory of Wightman functions, Commun. Math. Phys. {\bf 152}, 627--645 (1993).
\bibitem{Ste2} O. Steinmann, Perturbative quantum electrodynamics and axiomatic field theory, Springer Berlin/Heidelberg/N.Y., 2000.
\bibitem{Sug} M. Sugiura: Fourier series of smooth functions on compact Lie groups, Osaka 
J. Math. {\bf 8}, 33--47 (1971).
\bibitem{SW} R. F. Streater, A. S. Wightman:
PCT, spin, statistics and all that.
New York: Benjamin 1964.
\bibitem{SVW} A. Strohmaier, R. Verch, M. Wollenberg: Microlocal analysis of quantum fields on curved space-time: Analytic wavefront sets and Reeh-Schlieder theorems, Journ. Math. Phys. {\bf 43} No. {\bf 11},  5514--5530 (2002).
\bibitem{Tag} E. A. Tagirov: Consequences of field quantization in de Sitter type cosmological models, Ann. Phys. {\bf 76},
561--579 (1973).
\bibitem{Vil} N. J. Vilenkin: Special functions and the theory of group
representations. Translations of mathematical monographs, Vol. 22.
American mathematical society, Providence, R. I. 1968.
\bibitem{Wa} R. M. Wald: Quantum field theory in curved space-time and black hole thermodynamics, Chicago Univ. Press 1993.
\bibitem{YF} C. N. Yang, D. Feldman: {The S-matrix in the Heisenberg representation}, Phys. Rev. {\bf 79}, 972--987 (1950).
\end{thebibliography}
\end{document}